\documentclass[prl,twocolumn,superscriptaddress]{revtex4-1}
\usepackage{graphicx}
\usepackage{dcolumn}
\usepackage{bm}
\usepackage{amsmath}
\usepackage{color}
\usepackage{wrapfig}

\begin{document}

\title{Magnetic Purcell effect in nanophotonics}

\author{Denis~G.~Baranov}
\email[]{denisb@chalmers.se}
\affiliation{Department of Physics, Chalmers University of Technology, 412 96 Gothenburg, Sweden}
\affiliation{Moscow Institute of Physics and Technology, 9 Institutskiy per., Dolgoprudny 141700, Russia}

\author{Roman~S.~Savelev}
\affiliation{ITMO University, Saint Petersburg, Russia}

\author{Sergey~V.~Li}
\affiliation{ITMO University, Saint Petersburg, Russia}

\author{Alexander~E.~Krasnok}
\affiliation{ITMO University, Saint Petersburg, Russia}
\affiliation{Department of Electrical and Computer Engineering, The University of Texas at Austin, Austin, Texas 78712, USA}

\author{Andrea Al{\`u}}
\affiliation{Department of Electrical and Computer Engineering, The University of Texas at Austin, Austin, Texas 78712, USA}

\begin{abstract}
Tailoring of electromagnetic spontaneous emission predicted by E.~M.~Purcell more than 50 years ago has undoubtedly proven to be one of the most important effects in the rich areas of quantum optics and nanophotonics. Although during the past decades the research in this field has been focused on electric dipole emission, the recent progress in nanofabrication and study of magnetic quantum emitters, such as rare-earth ions, has stimulated the investigation of the magnetic side of spontaneous emission. Here, we review the state-of-the-art advances in the field of spontaneous emission enhancement of magnetic dipole quantum emitters with the use of various nanophotonics systems. We provide the general theory describing the Purcell effect of magnetic emitters, overview realizations of specific nanophotonics structures allowing for the enhanced magnetic dipole spontaneous emission, and give an outlook on the challenges in this field, which remain open to future research.
\end{abstract}

\maketitle
\section{Introduction}

\begin{figure}[!b]
\centering
\includegraphics[width=0.5\textwidth]{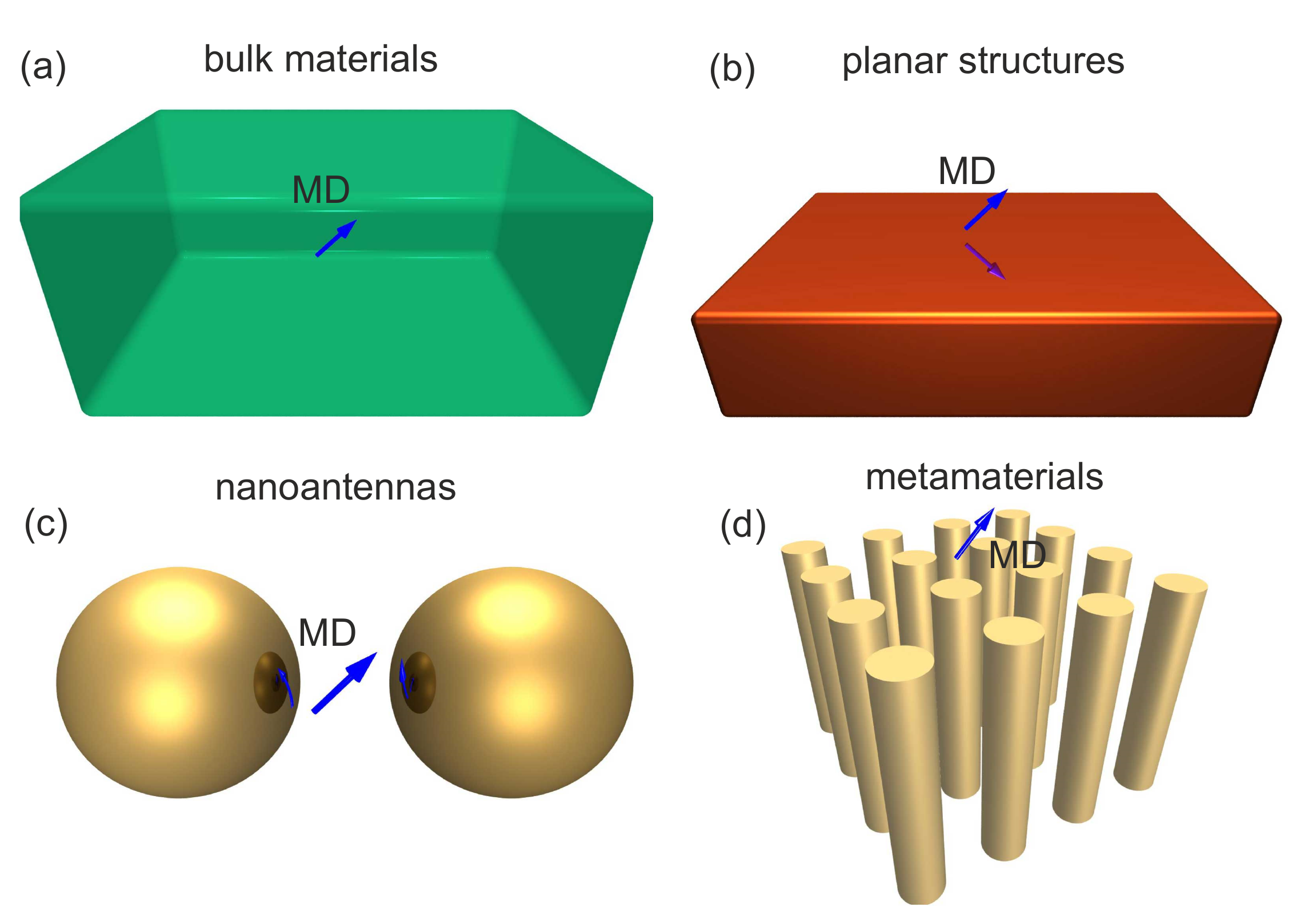}
\caption{Various nanophotonic structures that are used for enhancement of the magnetic dipole spontaneous emission rate: (a) bulk materials, (b) planar structures, (c) optical nanoantennas and (d) metamaterials.}
\label{Structures}
\end{figure}

Spontaneous emission of a quantum source into free space is a key phenomenon across the whole area of quantum optics. This phenomenon, described in the classical textbooks~\cite{Scully, Allen}, underlies the generation of almost all visible radiation -- from tiny light emitting diodes to the distant stars. Due to the very general origin of this effect, it is ubiquitous to a wide range of quantum sources, including atomic and molecular transitions, quantum dots and quantum wells, defect centers in nanocrystals, superconductor qubits and others. An important feature of the spontaneous emission discovered by E.~M.~Purcell in 1946 is that its rate can be either increased or decreased by varying an electromagnetic environment of the source~\cite{Purrel1946PR}. Since then the modification of the spontaneous emission rate of a quantum source induced by its interaction with the environment has been referred to as the \textit{Purcell effect}. Although this effect was originally discussed in the context of nuclear magnetic resonance, it applies to the broad range of quantum sources, that demonstrate spontaneous decay from the excited state~\cite{Krasnok_SR2015, Pelton}.

Nowadays, the Purcell effect is one of the cornerstones of quantum optics and nanophotonics. Strong Purcell effect is usually accomplished by placing emitters in the microcavities~\cite{Haroche_89, Slusher_1993, Vahala_2003}, hotspots of resonant nanoantennas~\cite{HulstNP2011, Brongersma_2010, NovotnyAOP, HechtNL2009, LarocheMarquierPRL2010, Acuna2012, Kinkhabwala2009, Song}, photonic crystals~\cite{Gaponenko, Lodahl2004, Vuckovic2005,Fujita1296, Noda2007} or metamaterials~\cite{Poddubny2013, Narimanov}. At optical frequencies, measurements of the spontaneous emission rate are usually performed via recording the time-resolved photoluminescence signal from quantum emitters excited in pulsed regime~\cite{Lodahl2004, Russell2012, Narimanov}. In the microwave range, in contrast, the Purcell effect can be observed as the enhancement of power radiated by a microwave antenna in the stationary regime~\cite{Krasnok_SR2015}. From the practical point of view, enhancement of the spontaneous emission rate is desired for such important applications as efficient laser operation~\cite{Scully}, single-photon sources~\cite{Arcari2014, Lodahl2015, Rao2007, Hoang2016, Portalupi2015}, fluorescent microscopy and nanoscale imaging~\cite{SandoghdarN2000, Acuna2012, Frimmer2011, Ropp2013, Beams2013}, biological studies~\cite{Ramamurthy2016PC}, and spectroscopy~\cite{Lakowicz2008}.

Up to date, the emphasis has been made on the investigation of the spontaneous decay of \textit{electric dipole} (ED) transitions, because the strength of ED transitions in typical optical quantum sources is orders of magnitude greater than that of \textit{magnetic dipole} (MD) transitions~\cite{Landau}. This difference is the reason why the permeability of natural materials is close to 1 in the visible range~\cite{Merlin}, and it also determines the difficulties of detecting and controlling the magnetic field at optical frequencies~\cite{Kuipers}. Nevertheless, certain quantum emitters, such as \textit{rare-earth ions}~\cite{Carnall1968, Judd1962, Ofelt1962} and \textit{semiconductor quantum dots}~\cite{Zurita-Sanchez2002}, possess prominent MD transitions whose strength is comparable or even greater than the competing ED ones. Growing interest of researchers in such emitters poses a challenging quest for nanostructures which can enable an enhanced interaction of light with MD quantum emitters and potentially lead to novel optical devices fully exploiting the magnetic nature of light.

\begin{figure*}
\centering
\includegraphics[width=0.99\textwidth]{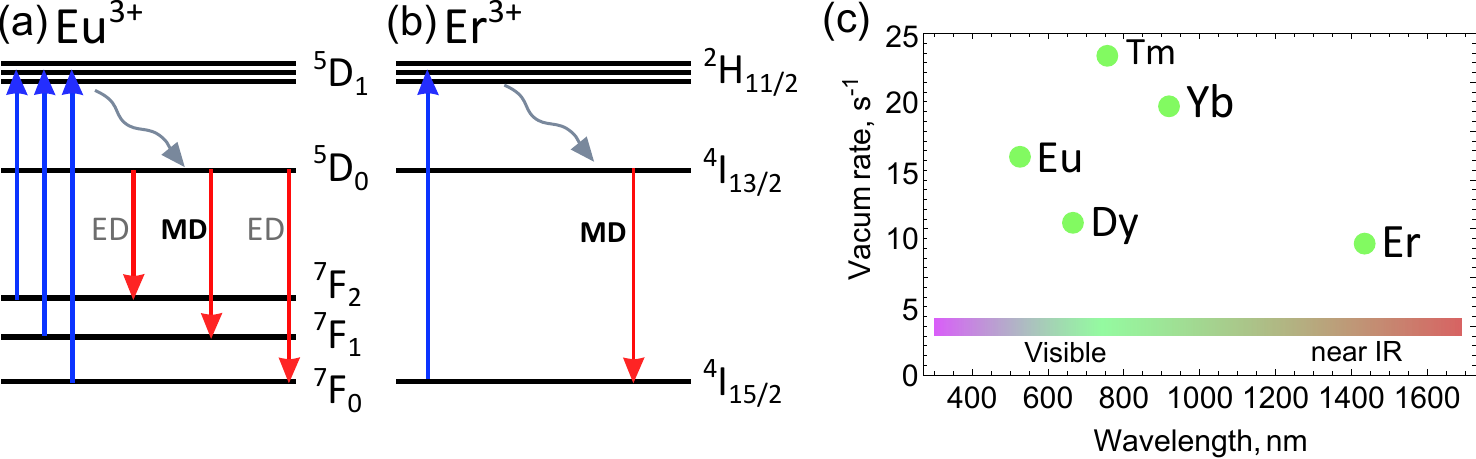}
\caption{The energy levels structure of the most frequently used MD optical emitters: (a) Eu$^{3+}$ ions and (b) Er$^{3+}$ ions. Blue arrows show the most frequently used excitation schemes. For Eu$^{3+}$, the strongest MD transition and several ED transitions with the shared upper state are shown. (c) Vacuum emission rates of the most representative rare-earth ions in the visible and near-IR regions. The data on the vacuum emission rates is adopted from Ref.~\cite{Dodson2012}.}
\label{spectrum}
\end{figure*}

There are a number of excellent review papers devoted to discussion of the spontaneous emission enhancement from ED transitions (see, e.g., Refs.~\cite{Pelton, Agio2012}). However, a comprehensive work summarizing the results on the Purcell effect for MD emitters, which would be highly valuable for researchers, is missing. In this paper, we are willing to review the state-of-the-art advances in the field of spontaneous emission enhancement of MD quantum emitters with the use of various nanophotonic systems including bulk materials, planar structures, optical nanoantennas, and metamaterials (see Fig.~\ref{Structures}). We provide the general theory describing Purcell effect of magnetic emitters, present an overview of available magnetic quantum sources and realizations of specific nanophotonic structures allowing for enhanced MD spontaneous emission, and give an outlook on the challenges in this field that remain open.

\section{Magnetic dipole optical emitters}

The relative weakness of the MD transitions can be understood in the picture of the Bohr atom model. A characteristic value of an atom ED moment is $d_0 \simeq {a_0}q$, where $a_0$ is the Bohr radius, and $q$ is the elementary charge. At the same time, the MD moment is of the order of Bohr magneton, $\mu_{\rm B} = \hbar e/2m$, where $m$ is the electron mass. Recalling that the electric ($E$) and magnetic ($B$) fields of a plane wave in free space are related via $B=E/c$ with $c$ being the speed of light  (we use the SI units throughout the paper), we conclude that the ratio of interaction constants of the MD and ED transitions with a plane electromagnetic wave is estimated as $1/\alpha \approx 137$. 
Nevertheless, there are certain materials that exhibit strong MD transitions in the optical range. Specifically, the fluorescence emission character of the rare-earth ions is naturally multipolar due to their unique 4$f$ orbitals~\cite{Judd1962}. Intensities of intra-4$f^n$ optical transitions of trivalent lanthanides were calculated in 1962, showing that the MD contributions in these transitions can be comparable to the contribution of ED transitions~\cite{Judd1962, Ofelt1962}. This theoretical prediction was confirmed experimentally in the late 1960s by Carnall et. al.~\cite{Carnall65, Carnall1968}.

There are several comprehensive review articles on the optical properties of rare-earth ions and ions-doped nanocrystals, methods of their synthesis, and practical applications~\cite{CondMat06, AccountsChem, Bednarkiewicz_12, Zhao_2013, Chen_2013}. Here we briefly discuss the most common materials employed for the magnetic Purcell effect applications. In most cases, these are insulators and semiconductors doped with trivalent rare earth ions~\cite{Kenyon_2002, Vetrone_2012}. The reason for the use of ions, rather than neutral atoms, is that during the doping process charged ions can be accelerated and directed along a specific direction, whereas it is not possible for neutral atoms. In addition, 
lanthanides are naturally occurring in their oxidized +3 form, which is the most stable one.

Historically, the emission of MD transition in optics was firstly observed around 1940 in the wide-angle interference experiments with europium ions (Eu$^{3+}$)~\cite{Deutschbein, Freed1941}, and since then the Eu$^{3+}$ complexes are the most common material for studying emission properties of MD sources in various environments. Besides the $^5D_0\rightarrow{^7}F_1$ MD transition, the emission spectrum of Eu$^{3+}$ ions also features several ED transitions from the same $^5D_0$ excited level to other states $^7F_0$, $^7F_2$ [see Fig.~\ref{spectrum}(a)]. The presence of several transitions with the shared upper state makes their rates dependent on each other, thus enabling an additional control over the fluorescence emission.

Another well-known MD source is erbium ions Er$^{3+}$ with the corresponding transition $^4I_{13/2}\rightarrow{^4}I_{15/2}$ lying in the telecommunication band at $\approx 1.55~\mu$m [see Fig.~\ref{spectrum}(b)]. The $^4I_{13/2}\rightarrow{^4}I_{15/2}$ transition of an erbium ion in vacuum is MD allowed while ED forbidden~\cite{Thommen:06}. However, when placed in a crystal lattice, degeneracy of the $4f$-states may be lifted by the Stark effect leading to mixed ED/MD character of the $^4I_{13/2}\rightarrow{^4}I_{15/2}$ transition~\cite{Judd1962, Thommen:06}.
Attractive wavelength of the $^4I_{13/2}\rightarrow{^4}I_{15/2}$ transition already made this element very important for optical telecommunication~\cite{Digonnet}. For instance, the erbium doped fiber amplifiers are vital for restoring the level of optical signals and allowing their propagation over very long distances. Handling the fluorescence emission by nanostructuring the environment makes erbium even more attractive for modern applications.

Other promising MD transitions in all trivalent lanthanides in ultraviolet, visible and near-infrared frequency ranges were identified quite recently in the systematic rigorous search performed by Dodson and Zia~\cite{Dodson2012}. In this study, comparing to the celebrated papers by Carnall et al~\cite{Carnall65, Carnall1968}, the possibility of transition between two excited states was also considered. Moreover, authors employed more complex model for the free ion Hamiltonian, including not only the electrostatic and spin-orbit interactions but also the two-body, three-body, spin-spin, spin-other-orbit, and electrostatically correlated spin-orbit interactions. Authors revealed several prominent MD transitions with significant oscillator strengths that can have some practical interest, Fig.~\ref{spectrum}(c).

In the context of the MD emission, an equally important issue is the problem of selective excitation of MD transitions. Since the strength of interaction of an ED transition with an incident electromagnetic field is dictated by the term $-({\mathbf{\hat d}} \cdot {\mathbf{E}})$ in the Hamiltonian with ${\mathbf{\hat d}}$ being the electric dipole moment operator, a zero of electric field is required for selective MD transition excitation. The most intuitive way to achieve this is to place a thin layer of quantum emitters in the node of a standing wave exhibiting a zero of the electric field and a maximum of the magnetic field~\cite{Cohen, Zheludev}.

Another elegant approach to this issue was employed in Ref.~\cite{Kasperczyk2015} with the use of an azimuthally polarized excitation beam, which is characterized by a zero of the electric field at its center and at the same time by a maximum of the magnetic field. In the experiment, Eu$^{3+}$-doped nanocrystals were pumped by an azimuthally polarized beam at the excitation wavelength of 527.5~nm ($^7$F$_0$$\rightarrow$$^5$D$_1$ MD transition). The resulting fluorescence map revealed a maximum of fluorescence intensity on the beam axis, where the electric field is zero. In turn, when the nanocrystal was excited at the wavelength of 532~nm ($^7$F$_1$$\rightarrow$$^5$D$_1$ ED transition), the fluorescence map demonstrated a ring-shape distribution of intensity, originating from spatial electric field distribution of an azimuthally polarized beam. This observation directly indicated selective MD transition excitation. A similar approach has been taken in numerical simulations in Ref.~\cite{Bermel} for simultaneous enhancement of MD excitation rate and suppression of ED emission. Specifically, enhanced excitation rate stemmed from the illumination of a doped nanocrystal with an azimuthally polarized beam, while suppression of ED emission was achieved with a special dielectric nanorod structure.

\section{Theoretical background}
\subsection{Electric dipole emission}

Owing to the universal nature of spontaneous decay in various systems, let us begin by recalling the general theory describing spontaneous emission of an electric dipole quantum emitter. The simplest model of a quantum emitter, yet reflecting many phenomena inherent to the real optical emitters, is a two-level system (TLS) with the ground $\left| g \right\rangle$ and $\left| e \right\rangle$ excited energy states separated by the transition energy $\hbar \omega_0$, where $\hbar$ is the reduced Planck constant. The transition process is characterized by the transition dipole moment matrix element ${{\mathbf{d}}_{eg}} = \left\langle e \right|q{\mathbf{\hat r}}\left| g \right\rangle$. A fundamental property of such system is that, once prepared in the excited state, it relaxes to the ground state being accompanied by the emission of a photon with energy $\hbar \omega_0$. Calculation of this process requires accounting for the interaction of a TLS with the continuum of electromagnetic modes of the free space. Such calculations have been performed by Weisskopf and Wigner in Ref.~\cite{Weisskopf}. They have demonstrated that the population of the excited state of the TLS located in free space exponentially decays with the rate $\gamma_0$:
\begin{equation}
{\gamma _0} = \frac{{\omega _0^3}}{{3\pi \hbar {\varepsilon _0}{c^3}}}{\left| {\mathbf{d_{eg}}} \right|^2},
\label{eq1}
\end{equation}
where $\varepsilon_0$ is the vacuum permittivity.

When the TLS is placed in a specific environment, its spontaneous emission rate changes~\cite{Abajo}. With the use of the Fermi golden rule one finds the modified emission rate~\cite{Scully}:
\begin{equation}
\gamma = \frac{{\pi {\omega _0}}} {{\hbar {\varepsilon _0}}}{\left| {\mathbf{d_{eg}}} \right|^2}{\rho _{\mathbf{n}}}\left( {{\mathbf{r_0}},{\omega _0}} \right),
\label{eq2}
\end{equation}
where ${\rho _{\mathbf{n}}}\left( {{\mathbf{r_0}},{\omega _0}} \right)$ is the local density of states (LDOS) of electromagnetic field at the TLS position $\mathbf{r_0}$:
\begin{equation}
{\rho _{\mathbf{n}}}\left( {{{\mathbf{r}}_0},{\omega _0}} \right) = \sum\limits_k {\left[ {{\mathbf{n}} \cdot {{\mathbf{e}}_k}({{\mathbf{r}}_0}) \otimes {\mathbf{e}}_k^*({{\mathbf{r}}_0}) \cdot {\mathbf{n}}} \right]\delta \left( {{\omega _k} - {\omega _0}} \right)} .
\label{eq3}
\end{equation}
Here, the summation runs over all eigenmodes ${{{\mathbf{e}}_k}}$ of the system with eigenfrequencies $\omega_k$. The eigenmodes ${{{\mathbf{e}}_k}}$ are the solutions to the homogeneous wave equation normalized by the condition $\int_V {\varepsilon ({\mathbf{r}}){{\mathbf{e}}_i}({\mathbf{r}}) \cdot {{\mathbf{e}}_j}({\mathbf{r}}){d^3}{\mathbf{r}}}  = {\delta _{ij}}$ with $\varepsilon (\mathbf{r})$ being the relative permittivity of the environment. The unit vector $\mathbf{n}$ points in the direction of the TLS dipole moment $\mathbf{d}_{eg}$. By varying the LDOS one can efficiently modify the spontaneous emission rate of the TLS.

Expression~(\ref{eq2}) gives the correct solution to the spontaneous emission problem in the weak coupling regime, when the interaction constant between the emitter and the electromagnetic modes is smaller than the decay rate of the electromagnetic mode $\gamma_a$~\cite{Scully}. This scenario corresponds to the so-called \emph{Markovian dynamics}, when the system does not remember its evolution and results in the exponential decay. In the opposite case of a strong coupling, the system dynamics is non-Markovian, and Eq.~(\ref{eq2}) can not be applied for the description of the spontaneous decay. In this case, the emitter may demonstrate non-exponential decay and \textit{Rabi oscillations} between the excited and ground states~\cite{Lodahl, Vuckovic, Hoeppe}.

\begin{figure}[!t]
\centering
\includegraphics[width=0.5\textwidth]{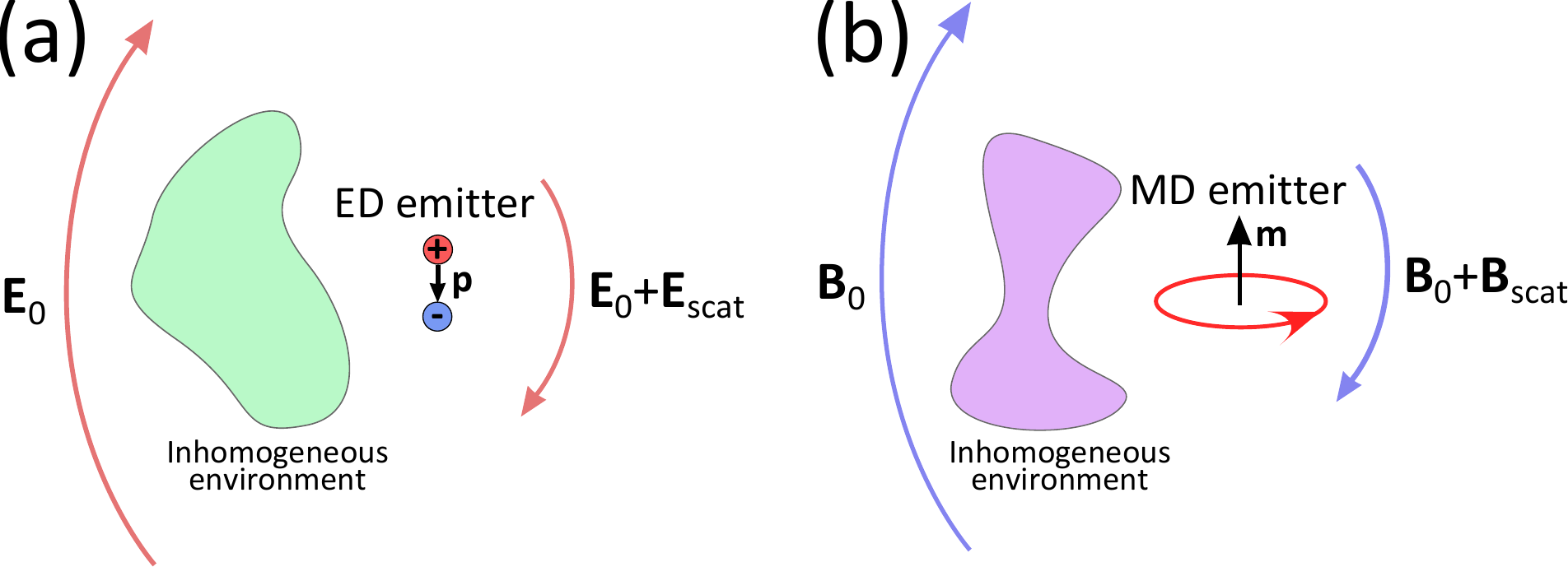}
\caption{Schematic illustration of the spontaneous emission enhancement realization by means of inhomogeneous environment for (a) electric and (b) magnetic dipole emitters.}
\label{concept}
\end{figure}

The electromagnetic LDOS in Eq.~(\ref{eq3}) may be calculated via expanding the dyadic Green tensor of an electric dipole source in the series of eigenmodes~\cite{Novotny}. For a closed and lossless cavity, this representation takes the form:
\begin{equation}
{\mathbf{G}}\left( {{\mathbf{r}},{\mathbf{r}}' ,\omega } \right) = \sum\limits_k {{c^2}\frac{{{\mathbf{e}}_k^*\left( {{\mathbf{r}} } \right) \otimes {{\mathbf{e}}_k}\left( {\mathbf{r}}' \right)}}{{\omega _k^2 - {\omega ^2}}}},
\label{eq4}
\end{equation}
where the Green tensor ${\mathbf{G}}\left( {\textbf{r},\textbf{r}',\omega_0 }\right)$ is the solution to the inhomogeneous wave equation:
\begin{equation}
\nabla \times \nabla \times {\mathbf{G}}\left( {{\mathbf{r}},{\mathbf{r}}',{\omega _0}} \right) - \varepsilon \left( {\mathbf{r}} \right)\frac{{\omega _0^2}}{{{c^2}}}{\mathbf{G}}\left( {{\mathbf{r}},{\mathbf{r}}',{\omega _0}} \right) = {\mathbf{I}}\delta \left( {{\mathbf{r}} - {\mathbf{r}}'} \right).
\label{eq5}
\end{equation}
with $\mathbf{I}$ being the unit dyad. The eigenmodes expansion~(\ref{eq4}) allows one to represent the LDOS as: ${\rho _{\mathbf{n}}}\left( {{{\mathbf{r}}_0},{\omega _0}} \right) = \frac{{2{\omega _0}}}{{\pi {c^2}}}{\mathbf{n}} \cdot \operatorname{Im}{\mathbf{G}}\left({{{\mathbf{r}}_0},{{\mathbf{r}}_0},{\omega _0}} \right) \cdot {\mathbf{n}}$. The resulting expression for the modified spontaneous emission rate takes the form:
\begin{equation}
\gamma = \frac{{2\omega _0^2}}{{\hbar {\varepsilon _0}{c^2}}}{\left| {{{\mathbf{d}}_{eg}}} \right|^2}{\mathbf{n}} \cdot \operatorname{Im} {\mathbf{G}}\left( {{{\mathbf{r}}_0},{{\mathbf{r}}_0},{\omega _0}} \right) \cdot {\mathbf{n}}.
\label{eq7}
\end{equation}
It is convenient to characterize the spontaneous emission rate by a dimensionless quantity, the \textit{Purcell factor}, $F_{\rm P}$, which represents the ratio between the spontaneous emission rate of a TLS in the given environment to the spontaneous emission rate of the same TLS in free space $\gamma_0$:
\begin{equation}
{F_{\text{P}}} = \frac{\gamma }{{{\gamma _0}}} = \frac{{6\pi c}}{{{\omega _0}}}{\mathbf{n}} \cdot \operatorname{Im} {\mathbf{G}}\left( {{{\mathbf{r}}_0},{{\mathbf{r}}_0},{\omega _0}} \right) \cdot {\mathbf{n}}.
\label{eq8}
\end{equation}
Remarkably, the value of $F_{\rm P}$ (under the weak coupling regime) does not depend on the TLS dipole moment magnitude and is only defined by the electromagnetic properties of the environment, which is reflected in the imaginary part of the Green tensor.

For a general open or lossy cavity the eigenmode expansion~(\ref{eq4}) can not be used, because the basis of integrable eigenmodes $\mathbf{e}_k (\mathbf{r})$ can not be defined in a straightforward way~\cite{Kristensen2014,Ginzburg}. Nevertheless, the Green tensor approach established in Eq.~(\ref{eq7}) gives the correct result for the spontaneous emission rate of an ED emitter located in an open resonator~\cite{Ginzburg}.

Expression~(\ref{eq7}) allows one to calculate the spontaneous emission rate of a quantum emitter given the knowledge of a classical characteristic -- the Green tensor ${{\mathbf{G}}}\left({{\mathbf{r}},{\mathbf{r}}',\omega } \right)$. Besides, this expression offers an interpretation of the Purcell effect in the classical terms. Indeed, the acceleration of spontaneous emission from an ED emitter can be understood as the enhancement of work done by the electric field of a dipole on the oscillating electric current, Fig.~\ref{concept}(a). The work performed by the electric field $\mathbf{E}$ of an oscillating dipole ${\mathbf{d}}{e^{ - i{\omega _0}t}}$ is given by
\begin{equation}
P = \frac{\omega }{2}\operatorname{Im} \left[ {{{\mathbf{d}}^*} \cdot {\mathbf{E}}\left( {{{\mathbf{r}}_0}} \right)} \right] = {\mu _0}\frac{{{\omega ^3}}}{2}{\left| {\mathbf{d}} \right|^2}{\mathbf{n}} \cdot \operatorname{Im} {\mathbf{G}}\left( {{{\mathbf{r}}_0},{{\mathbf{r}}_0},{\omega _0}} \right) \cdot {\mathbf{n}},
\label{eq9}
\end{equation}
where $\mu_0$ is the vacuum permeability. Dividing this quantity by the power ${P_0} = \tfrac{{{\omega_0 ^4}}}{{12\pi {\varepsilon _0}{c^3}}}{\left| {\mathbf{d}} \right|^2}$ emitted by the same dipole into free space, we obtain expression identical to the Purcell factor (\ref{eq8}).

The total decay rate $\gamma$ introduced by Eq.~(\ref{eq7}) accounts for both radiative and non-radiative processes~\cite{Ginzburg}. In most practical cases the enhancement of power emitted to the far field is of specific interest, thus an appropriate figure of merit is the radiative Purcell factor $F_{\text{P}}^{(r)} = {\gamma _r}/{\gamma _0}$. Here, the radiative decay rate $\gamma_r$ can be calculated by integrating the Poynting vector over a surface enclosing the dipole and its environment. In order to characterize the fraction of energy emitted into photons one also defines the \textit{quantum yield} defined as
\begin{equation}
\eta = \frac{{{\gamma _{\text{r}}}}}{{{\gamma _{\text{r}}} + {\gamma _{{\text{nr}}}} + {\gamma _{\rm int }}}},
\end{equation}
where ${{\gamma _{{\text{nr}}}}}$ is the decay rate due to the electromagnetic loss in the environment and ${{\gamma _{{\text{int}}}}}$ is the rate of the intrinsic non-radiative decay that occurs even when an isolated emitter is placed in free space. By introducing the vacuum quantum yield $\eta_0$ which reflects the fraction of energy emitted by an isolated TLS, the resulting quantum yield of an emitter in the environment takes the form:
\begin{equation}
\eta = {\eta _0}\frac{{F_{\text{P}}^{(r)}}}{{{\eta _0}{F_{\text{P}}} + (1 - {\eta _0})}}.
\end{equation}

\subsection{Single-mode approximation}
The expression (\ref{eq8}) for the Purcell factor may be simplified if the main contribution to the Green tensor at the emitter frequency is provided by a certain electromagnetic mode. In this specific case, one may use a single-mode approximation and write the Green tensor as:
\begin{equation}
{\mathbf{G}}\left( {{\mathbf{r}},{\mathbf{r'}},{\omega _0}} \right) \approx {c^2}\frac{{{\mathbf{e}}_k^*\left( {{\mathbf{r}}} \right) \otimes {{\mathbf{e}}_k}\left( {\mathbf{r}'} \right)}}{{\omega _k^2 - \omega _0^2 - 2i{\gamma _k}\omega }},
\label{eq10}
\end{equation}
where ${\gamma _k}$ is the damping rate of the $k$-th resonant eigenmode ${{\mathbf{e}}_k}\left( {\mathbf{r}} \right)$. Assuming that the emitter, whose frequency matches that of the cavity mode, is located at the maximum of the eigenmode electric field and its dipole moment is collinear with the electric field polarization, we come to the~well-known expression for the single-mode Purcell factor~\cite{Purrel1946PR}:
\begin{equation}
F_{\rm P} = \frac{3}
{{4{\pi ^2}}}{\lambda ^3}\frac{Q}{V},
\label{eq11}
\end{equation}
where $Q = {\omega _k}/2{\gamma _k}$ is the mode quality factor, $\lambda$ is the free space wavelength and $V$ is the \emph{mode volume} defined according to $V = \tfrac{1}{{{{\left| {{{\mathbf{e}}_k}\left( {{{\mathbf{r}}_0}} \right) \cdot {\mathbf{n}}} \right|}^2}}}$.

The single-mode approximation does not always give the accurate result, in particular, for an emitter placed close to a plasmonic nanoantenna~\cite{Koenderink2010, Sauvan2013}. Moreover, for an open or dissipative nanostructure, the definition of the mode volume is nontrivial. 
The quasinormal modes of such structures have complex-valued eigenfrequencies, resulting in exponential divergence of the electromagnetic field at large distances from the system~\cite{Kristensen2014}, so that the normalization condition $\int_V {\varepsilon ({\mathbf{r}}){{\mathbf{e}}_i}({\mathbf{r}}) \cdot {{\mathbf{e}}_j}({\mathbf{r}}){d^3}{\mathbf{r}}}  = {\delta _{ij}}$ cannot be satisfied.
Nevertheless, certain approaches for calculation of the mode volume for open nanostructures have been suggested that rely on different normalization procedures~\cite{Kristensen2014, Sauvan2013, HughesPRA, KristensenOL}.

\subsection{Magnetic dipole emission}
The above calculations may be mapped in a straightforward fashion to the description of the spontaneous emission from a MD emitter. The spontaneous emission rate $\gamma^{(m)}$ of a TLS with the transition MD moment ${{\mathbf{m}}_{eg}} = {\mu _{\text{B}}}\left\langle e \right|{\mathbf{\hat L}} + 2{\mathbf{\hat S}}\left| g \right\rangle$, where ${\mathbf{\hat L}}$ and ${\mathbf{\hat S}}$ are the operators of the orbital and spin momentum, respectively, is given by the Fermi rule:
\begin{equation}
{\gamma ^{(m)}} = \frac{{\pi {\omega _0}}}{\hbar }{\mu _0}{\left| {{{\mathbf{m}}_{eg}}} \right|^2}\rho _{\mathbf{n}}^{(m)}\left( {{{\mathbf{r}}_0},{\omega _0}} \right),
\label{eq14}
\end{equation}
where $\rho _{\mathbf{n}}^{(m)}\left( {{\mathbf{r}},{\omega _0}} \right)$ is the \emph{magnetic} local density of states for the given environment:
\begin{equation}
\rho _{\mathbf{n}}^{(m)}\left( {{{\mathbf{r}}_0},\omega } \right) = \frac{{2{\omega _0}}}{{\pi {c^2}}}{\mathbf{n}} \cdot \operatorname{Im} {{\mathbf{G}}^{(m)}}\left( {{{\mathbf{r}}_0},{{\mathbf{r}}_0},{\omega _0}} \right) \cdot {\mathbf{n}}.
\label{eq15}
\end{equation}
Here, ${{\mathbf{G}}^{(m)}}\left( {{\mathbf{r}},{\mathbf{r'}},{\omega _0}} \right)$ is the Green tensor of a magnetic dipole, which connects the magnetic field at the position $\mathbf{r}$ to the magnetic dipole located at ${{\mathbf{r'}}}$ via ${\mathbf{H}}\left( {\mathbf{r}} \right) = k_0^2{{\mathbf{G}}^m}\left( {{\mathbf{r}},{{\mathbf{r}}_0},{\omega _0}} \right){\mathbf{m}}$ with $k_0=\omega_0/c$.
Using Eq.~(\ref{eq15}) we arrive at the expression relating the MD spontaneous emission rate to the magnetic Green tensor:
\begin{equation}
{\gamma ^{(m)}} = \frac{{2{\mu _0}\omega _0^2}}{{\hbar {c^2}}}{\left| {{{\mathbf{m}}_{eg}}} \right|^2}{\mathbf{n}} \cdot \operatorname{Im} {{\mathbf{G}}^{(m)}}\left( {{{\mathbf{r}}_0},{{\mathbf{r}}_0},{\omega _0}} \right) \cdot {\mathbf{n}}.
\label{eq16}
\end{equation}
Finally, the magnetic Purcell factor is defined as the enhancement of a MD emitter decay rate with respect to the vacuum value $\gamma _0^{(m)} = \frac{{\omega _0^3}}{{3\pi \hbar {c^3}}}{\mu _0}{\left| {{{\mathbf{m}}_{eg}}} \right|^2}$, which is obtained from Eq.~(\ref{eq14}) noting that the electric and magnetic LDOS are equal in free space due to the symmetry of Maxwell's equations~\cite{Carminati2003}:
\begin{equation}
F_{\text{P}}^{(m)} = \frac{{6\pi c}}{{{\omega _0}}}{\mathbf{n}} \cdot \operatorname{Im} {{\mathbf{G}}^{(m)}}\left( {{{\mathbf{r}}_0},{{\mathbf{r}}_0},{\omega _0}} \right) \cdot {\mathbf{n}}.
\label{eq17}
\end{equation}

As for the case of electric dipole, the expression~(\ref{eq17}) may be obtained from a classical argument by considering the work $P^{(m)}$ performed by the magnetic field of a classical magnetic dipole $\mathbf{m}$ on the oscillating magnetic current, Fig.~\ref{concept}(b):
\begin{equation}
\begin{gathered}
  {P^{(m)}} = \frac{\omega }{2}\operatorname{Im} \left[ {{{\mathbf{m}}^*} \cdot {\mathbf{B}}\left( {{{\mathbf{r}}_0}} \right)} \right] \hfill \\
   = \frac{{{\mu _0}}}{{{c^2}}}\frac{{{\omega ^3}}}{2}{\left| {\mathbf{m}} \right|^2}{\mathbf{n}} \cdot \operatorname{Im} {{\mathbf{G}}^{(m)}}\left( {{{\mathbf{r}}_0},{{\mathbf{r}}_0},{\omega _0}} \right) \cdot {\mathbf{n}} \hfill.
\end{gathered} 
\label{eq18}
\end{equation}

One can also calculate the magnetic Purcell factor in the single-mode approximation if magnetic response of a cavity at the emitter frequency is dominated by a specific eigenmode. The resulting expression for $F_{\rm P}$ is identical to Eq.~(\ref{eq11}) up to substitution of the electric mode volume $V$ with the magnetic mode volume 
${V^{(m)}} = \tfrac{1}{{{{\left| {{{\mathbf{B}}_k}\left( {{{\mathbf{r}}_0}} \right) \cdot {\mathbf{n}}} \right|}^2}}}$~\cite{Zambrana-Puyalto2015},
 where $ \mathbf{B}_k ( \mathbf{r}_0)$ is the magnetic field of the normalized eigenmode at the emitter location.

\subsection{Chiral emitters}
Within the context of MD spontaneous emission, it is instructive to briefly discuss the Purcell effect for chiral emitters. The electronic transition of a chiral molecule is characterized by its electric $\mathbf{d}$ and magnetic $\mathbf{m}=\pm i \xi \mathbf{d}$ dipole moments, where the plus and minus signs correspond to the right and left \emph{enantiomers} of the molecule, and $\xi$ is a real value depending on the internal structure of the emitter. Because of the different relative orientation of dipole moments in the two enantiomers, one may expect that the spontaneous emission rates of the right and left molecules can be controlled by chiral properties of the environment.

The power performed by the field of a chiral molecule is given by
\begin{equation}
{P^{(ch)}} = \frac{\omega }
{2}\operatorname{Im} \left( {{{\mathbf{d}}^*}{\mathbf{E}}\left( {{{\mathbf{r}}_0}} \right) + {{\mathbf{m}}^*}{\mathbf{B}}\left( {{{\mathbf{r}}_0}} \right)} \right).
\label{eqChiral0}
\end{equation}
Now, however, induced electric and magnetic fields should be related to both electric and magnetic dipole moments via
\begin{equation}
\begin{gathered}
  {\mathbf{E}}\left( {\mathbf{r}} \right) = \frac{{k_0^2}}{{{\varepsilon _0}}}{\mathbf{G}}\left( {{\mathbf{r}},{{\mathbf{r}}_0},{\omega _0}} \right){\mathbf{d}} + \frac{{i{k_0}}}{{c{\varepsilon _0}}}\nabla  \times {{\mathbf{G}}^{(m)}}\left( {{\mathbf{r}},{{\mathbf{r}}_0},{\omega _0}} \right){\mathbf{m}}, \hfill \\
  {\mathbf{H}}\left( {\mathbf{r}} \right) = k_0^2{{\mathbf{G}}^{(m)}}\left( {{\mathbf{r}},{{\mathbf{r}}_0},{\omega _0}} \right){\mathbf{m}} + \frac{{{k_0}c}}{i}\nabla  \times {\mathbf{G}}\left( {{\mathbf{r}},{{\mathbf{r}}_0},{\omega _0}} \right){\mathbf{d}}. \hfill \\ 
\end{gathered} 
\end{equation}
Due to the cross-coupling of electric and magnetic moments in Eq.~(\ref{eqChiral0}) the total power is not a simple sum of expressions~(\ref{eq9}) and~(\ref{eq16}). Now, the total power contains a term proportional to the multiple of the two dipole moments $\operatorname{Im} \left( {{{\mathbf{d}}^*}{\mathbf{m}}} \right)$. Consequently, the spontaneous emission rate of chiral emitters may be controlled via chirality of the environment.

Overall, the expressions (\ref{eq14})--(\ref{eq17}) given above outline the general route towards enhanced magnetic spontaneous emission. It is analogous to that for ED emitters: one needs to \textit{engineer a nanostructure with the high magnetic local density of states}. Or, in the framework of classical electromagnetism, one needs a large imaginary part of the Green tensor at the position of the MD emitter.

\begin{figure}
\centering
\includegraphics[width=0.5\textwidth]{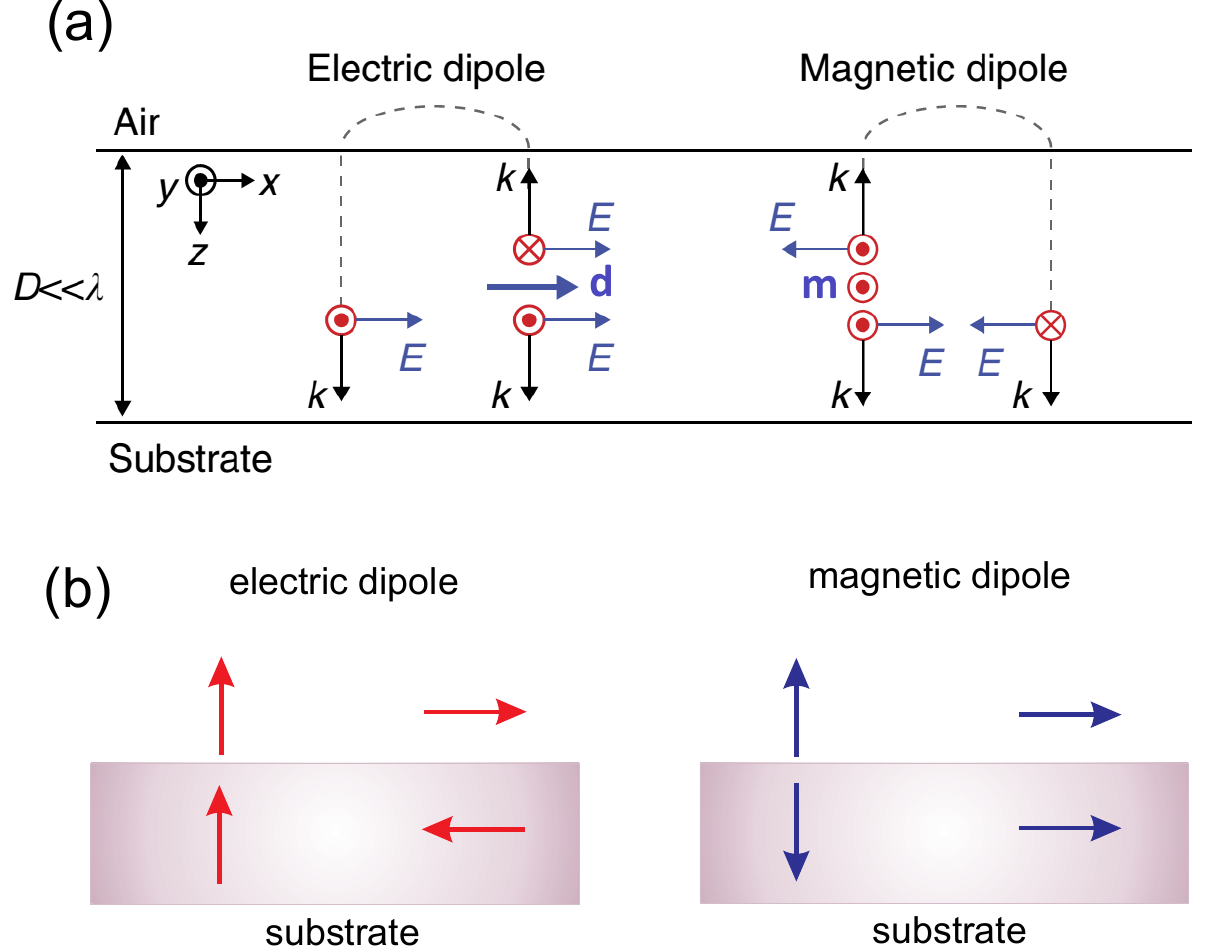}
\caption{(a) An example of constructive and destructive interference from the electric and magnetic dipoles embedded in a ultrathin layer near a substrate. Adapted from Ref.~\cite{ZiaNatComm2012}. (b)~Images of electric and magnetic dipoles induced in a metallic substrate.}
\label{dipoles}
\end{figure}

\section{Magnetic Purcell factor in bulk and planar structures}

\begin{figure*}
\centering
\includegraphics[width=0.99\textwidth]{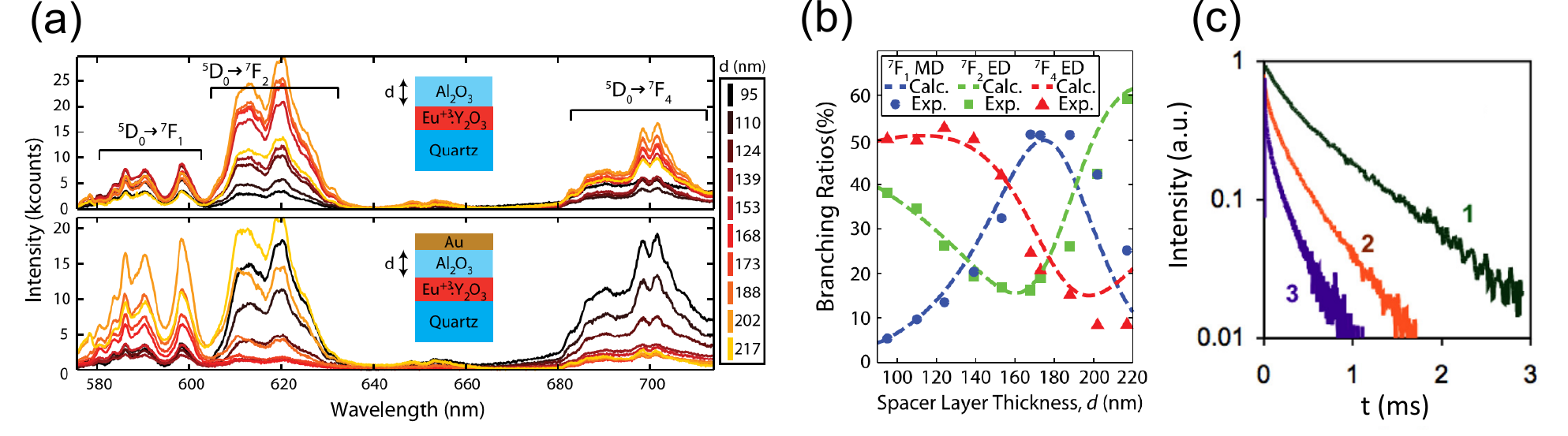}
\caption{Enhancement of spontaneous MD emission in planar structures. (a)~Emission spectra from uncoated (top) and gold-coated (bottom) Eu$^{3+}$-doped thin films for different thickness ($d$) of Al$_2$O$_3$ spacer layer. Three prominent peaks correspond to two ED transitions ($^5D_0\rightarrow{^7}F_2$ and $^5D_0\rightarrow{^7}F_4$) and to one MD transition ($^5D_0\rightarrow{^7}F_1$). (b)~Measured and calculated branching ratios $\beta_i$ (see the definition in text) of three different transitions of Eu$^{3+}$ ions for the geometry shown in (a) as a function of distance $d$ from the gold substrate. Adapted from Ref.~\cite{ZiaPRL2011}. (c)~Time-resolved emission intensity of Eu$^{3+}$ ions deposited on glass (1), gold (2) and gold nanostrips (3).  Adapted from Ref.~\cite{NoginovaOE2009}.}
\label{Zia}
\end{figure*}

Probably, the simplest way to modify the local density of optical states is to place an emitter inside a dielectric medium with refractive index $n$ different from unity.
For an ED emitter, the microscopic electric field at the position the dipole not only is modified through the Green tensor of a homogeneous medium but also is changed due to the \textit{local field effect}. The resulting modified spontaneous emission rate can be expressed as follows~\cite{Glauber}:
\begin{equation}
\gamma_{\rm ED} (n) = n L^2 {\gamma _0},
\label{eq:gamma_ED_bulk}
\end{equation}
where $\gamma_0$ is the spontaneous emission rate in vacuum, and $L$ is the local field factor~\cite{Jackson}, which differs for various models (for a review of the local field effects see, e.g., Ref.~\cite{Boyd}). For a MD emitter, however, the result is different, because the microscopic magnetic field does not depend on $n$~\cite{Jackson}, and the permeability of dielectric materials is usually equal to one. Consequently, the local field effects for the MD emitter are absent. Therefore the spontaneous emission rate is modified only through the magnetic LDOS which is proportional to the third power of $n$~\cite{Glauber}:
\begin{equation}
\gamma_{\rm MD} (n) = n^3 {\gamma _0}.
\label{eq:gamma_MD_bulk}
\end{equation}
Thus, unlike ED transitions, the emission rates of MD emitters are more sensitive to the host matrix refractive index. These theoretical predictions were experimentally confirmed in Ref.~\cite{RikkenPRL1995} through the measurements of spontaneous emission rates of ED and MD transitions in Eu$^{3+}$ complexes, placed in solvents with different refractive indices. 

In reality, placing a rare-earth ion in a crystal modifies the structure of its energy levels due to the Stark effect of the crystal static electric field. As a result, the ED transitions, being forbidden for an ion in free space, become ED allowed if ion resides in a nonsymmetric site of the lattice where the crystal field is non-zero.
Therefore, the rates of ED transitions can be substantially reduced by embedding ions in the centrosymmetric sites of the crystal lattice, while at the same time the rates of allowed MD transitions weakly depend on the symmetry of the environment~\cite{Atwood}. Comparison of the calculations made for MD transitions of Yb$^{3+}$ ions in different host materials in Ref.~\cite{Dodson2012} with the experimental measurements done in Ref.\cite{Krupke} revealed that up to 50\% of all decay processes can result in MD emission in centrosymmetric environment.

Another simple route to the magnetic (as well as the electric) LDOS  modification implies placing a MD emitter near the interface of two media, or, generally, in some layered structure. This approach seems to be more realistic in terms of possible practical applications, and consequently it is well-studied experimentally.
Radiation of a point dipole located near an interface between two media was first considered in the beginning of the 20th century by Arnold Sommerfeld~\cite{Sommerfeld}. To date, the general classical electrodynamics theory of a point dipole radiation in layered structures is well established for both ED and MD emitters. Corresponding analytical expressions, allowing to calculate spontaneous emission rates can be found e.g. in Refs.~\cite{LukoszJOSA1977-1, LukoszJOSA1977-2, LukoszJOSA1979} and textbook~\cite{Novotny}.

The different picture of the spontaneous emission of ED and MD emitters located near layered structures originates from the symmetry of the electromagnetic field emitted by electric and magnetic dipoles, respectively, Fig.~\ref{dipoles}(a). While the electric far field of an electric dipole is \emph{symmetric} under mirror transformation, the electric field of a magnetic dipole is \emph{antisymmetric}. This symmetry of electric field further leads to different interference of the electric field of a dipole itself and the field reflected from the interface between the emitters layer and air in the far field zone, as schematically depicted in Fig.~\ref{dipoles}(a). This difference may also be understood with the image dipoles concept, Fig.~\ref{dipoles}(b) -- emitter interferes with its image constructively or destructively, depending on the phase delay. It allows to distinguish the type of transition from direct measurements of spontaneous emission rate dependence on the distance between the emitter and the interface, and to employ planar dielectric and metallic structures for modification of the MD emitters spontaneous emission rate.

\begin{figure*}
\centering
\includegraphics[width=0.8\textwidth]{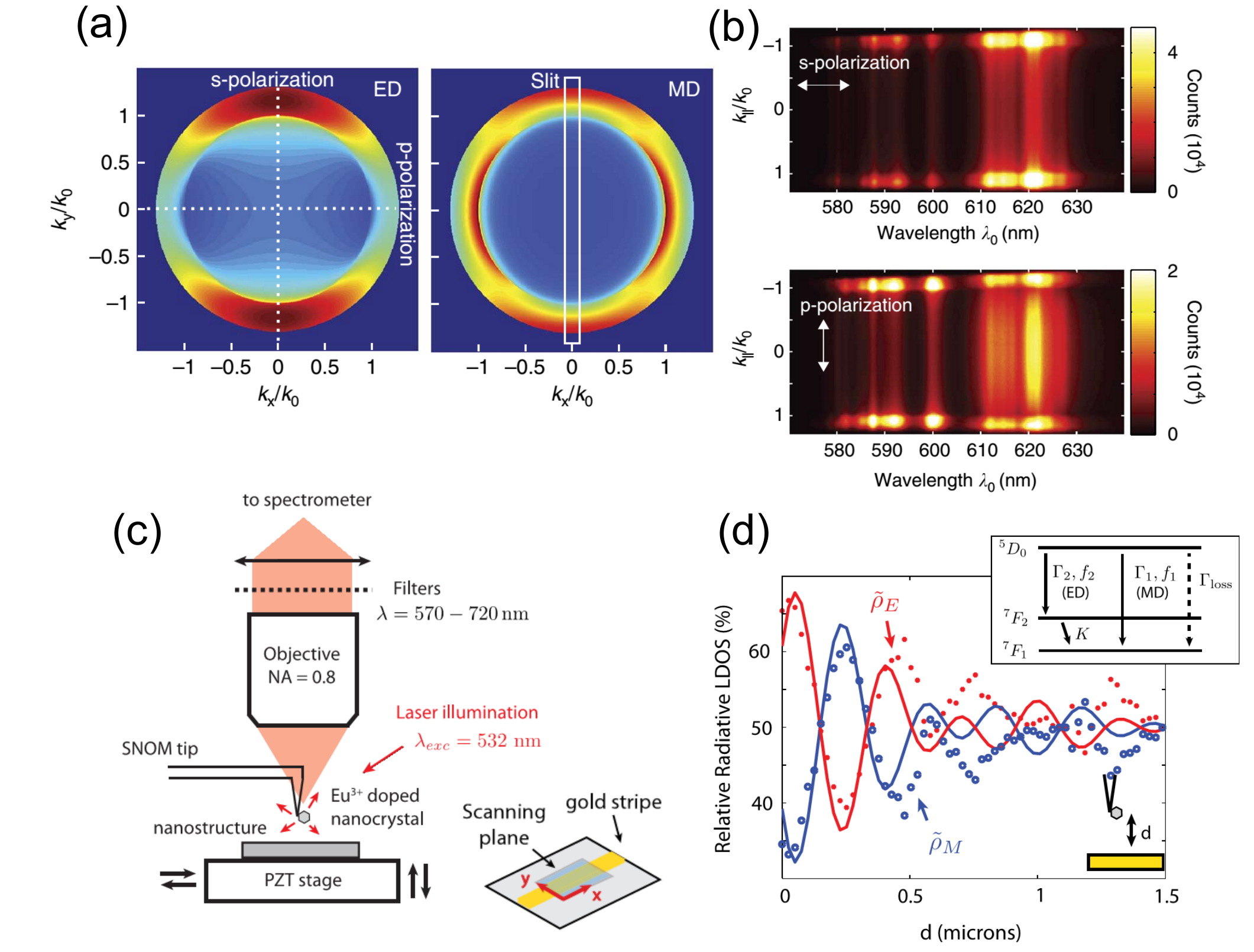}
\caption{Spectroscopic techniques for mapping ED and MD transitions. (a)~Calculated momentum spectra produced by isotropic ED and MD emitters embedded into a 21 nm thick dielectric substrate. The spectra are cut off at $k_{x,y}/k_0=1.3$ corresponding to the objective numerical aperture NA$=1.3$. (b)~Measured energy-momentum resolved s- and p-polarized fluorescence emission spectra of Eu$^{3+}$ doped film. Adapted from Ref.~\cite{ZiaNatComm2012}. (c)~Schematic of the setup for 3D mapping of the magnetic LDOS developed in Ref.~\cite{CarminatiPRL2014}. A Eu$^{3+}$ nanocrystal is attached to an optical near-field microscope. The sample can move in all three directions enabling 3D mapping of luminescence. (d) Measured electric and magnetic relative LDOS as a function of distance from the gold substrate. Adapted from Ref.~\cite{CarminatiPRL2014}.}
\label{spectr}
\end{figure*}

Experimental studies of fluorescence modification of the europium complexes in the ultrathin films (several nanometers) placed near the gold and dielectric substrates started more than 50 years ago. A summary of the results stemming from these experiments is given in details in Ref.~\cite{Drexhage1974} (see also refs therein). These experiments were focused on the study of modification of directivity patterns and spontaneous emission rates of Eu$^{3+}$ ions placed near the substrate at the different distances from the substrate. The distance was precisely controlled by changing the number of monomolecular layers of CdC$_{20}$, located between the substrate and the europium layer. The results clearly demonstrated the difference in the dependence of the fluorescence intensity on the distance for 592~nm peak (MD transition) and 612~nm (ED transition), which allowed to distinguish the nature of these transitions.

In the last few years, thanks to the advances in nanofabrication, a revival of interest in experiments related to the spontaneous emission of rare-earth ions in planar structures is observed~\cite{Karaveli2010, ZiaPRL2011, ZiaNatComm2012, ZiaACSNano2013, ZiaPRB2014, ZiaOME2014, HussainOE2014}. The emerging possibilities for improved control of lanthanide ions emission by simple use of various substrates may empower the modern capacity of nanooptics tools and facilitate the development of efficient magnetic field optical probes~\cite{NoginovaJAP2008, NoginovaOE2009, NiAPB2011}.
For instance, Noginova \textit{et al.} proposed for the first time that systems containing Eu$^{3+}$ ions can be used as spectroscopic tools for measurement of the magnetic optical fields at the nanoscale~\cite{NoginovaJAP2008}. It has been demonstrated that the study of the MD transition at $\approx$590~nm (see Fig.~\ref{spectrum}(a)) in comparison with the ED transition at $\approx$615~nm can reveal information on the relative distribution of magnetic and electric fields as well as effective permeability and permittivity. Today, this approach is widely used for microscopy of electric and magnetic LDOS at the nanometer scale.

Lanthanides are characterized by multilevel electronic structure, i.e. the upper excited state can relax to one of the several lower states. This establishes the competition between different transitions, so the suppression of the emission through one of the transitions enhances the emission through others. Such peculiarity was employed in study of the fluorescence properties of europium chelate in Refs.~\cite{Karaveli2010, ZiaPRL2011} (Fig.~\ref{Zia}(a)). In order to quantitatively determine the fraction of energy that is emitted through the MD transition of Eu$^{3+}$ ions, the authors introduced the branching ratio, which is defined as follows: ${\beta_{i}} = \int_{\lambda_i} {{I}(\lambda )d\lambda } /\int_{580}^{715} {I(\lambda )d\lambda }$, where subscript $i$ denotes one of the transitions $^7F_1$ (MD), $^7F_2$ (ED), and $^7F_4$ (ED), and $\lambda_i$ is the wavelength range of the $i$th transition emission (580-603~nm, 603-635~nm and 680-715~nm for $^7F_1$, $^7F_2$ and $^7F_4$, respectively). Experimentally measured dependence of branching ratio on the thickness of Al$_2$O$_3$ spacer layer $d$ [see inset in Fig.~\ref{Zia}(a)] is shown in Fig.~\ref{Zia}(b). By varying the value of $d$ authors managed to direct up to the 50\% of all emission through one of the possible transitions. In Ref.~\cite{Karaveli2010} in slightly different geometry only 25\% value of branching ratio was achieved for MD transition [Fig.~\ref{Zia}(b)], while MD enhancement factor was measured to be equal $\approx$ 4, as compared to the reference glass sample. It should be noted, that in these papers fluorescent ions were placed at the distances about 100~nm from the metal layer. In Ref.~\cite{HussainOE2014} it was shown that if emitters are placed in the very proximity (tens of nanometers and less) to a metal layer, the emission intensity could decrease to zero due to strong quenching of the fluorescence. Finally, direct time-resolved measurements of MD spontaneous emission were performed in Ref.~\cite{NoginovaOE2013} (Fig.~\ref{Zia}(c)) for Eu$^{3+}$ ions, demonstrating accelerated dynamics of the emission. Here, the green curve (1) corresponds to the the single Eu$^{3+}$ crystals, red curve (2) -- the glass substrate, and purple curve (3) -- Eu$^{3+}$ ions deposited on the mirror.

\begin{figure*}
\centering
\includegraphics[width=0.99\textwidth]{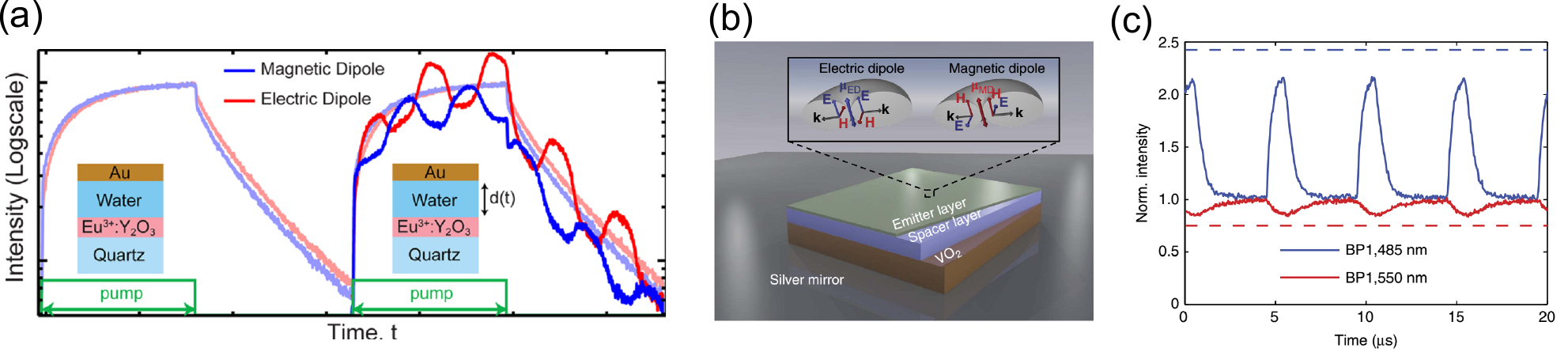}
\caption{Ultrafast modulation of spontaneous emission from rare-earth ions. (a)~Ultrafast modulation of spontaneous emission from ED and MD emitters achieved via a rapid change of thickness $d$ of the water layer between the gold mirror and layer with Eu$^{3+}$ ions. Adapted from Ref.~\cite{S.Karaveli2013}. (b)~Schematic of sub-lifetime dynamic control of spontaneous emission via phase-change material VO$_2$. (c)~Time-resolved normalized photoluminescence of the Er$^{3+}$ ions in the multilayer structure upon periodic VO$_2$ switching. Adapted from Ref.~\cite{ZiaNatComm2015}.}
\label{modulation}
\end{figure*}

On the basis of these results, an experimental method of energy-momentum spectroscopy that allows to directly distinguish different multipolar contributions into the total fluorescence has been developed~\cite{ZiaNatComm2012}. This method is based on the back-focal plane imaging~\cite{NovotnyJOSAb2004} with additional spectral analysis of the images. Typical back-focal plane images of Eu$^{3+}$ ions emission with bandpass filters for 605--635~nm and 585--595~nm are shown in Fig.~\ref{spectr}(a). Such images alone allow to determine the multipole nature of the transition. The additional spectral analysis provides back-focal plane spectra, shown in Fig.~\ref{spectr}(b) for europium ions. Such spectra allow to fully characterize all properties (i.e. wavelength, momentum and polarization) of optical mode into which the emitter radiates. Moreover, analytical calculation of the electric and magnetic LDOS in a given structure along with experimental measurement of counts originating from ED and MD transitions allows to obtain the intrinsic MD and ED spontaneous emission rates, which were shown to be actually comparable for Eu$^{3+}$ ions. Later, energy-momentum spectroscopy method along with the measurement of the time-dependent fluorescence intensity was also employed for characterization of the fluorescence of different rare-earth ions: Cr$^{3+}$ in magnum oxide~\cite{ZiaACSNano2013}, and Er$^{3+}$~\cite{ZiaPRB2014}, Dy$^{3+}$ è Tm$^{3+}$~\cite{ZiaOME2014} in yttrium oxide in the near-infrared frequency range.

Another method of 3D mapping of relative electric and magnetic LDOS has been reported in Ref.~\cite{CarminatiPRL2014}. Measurements of luminescence of a Eu$^{3+}$ doped nanocrystal placed in the vicinity of a gold mirror [see Fig.~\ref{spectr}(c)] allowed to extract the branching ratios of the two ED and one MD Eu$^{3+}$ transitions around 600~nm. These values can be further used to recover the relative contributions of electric and magnetic components at frequencies $\omega_1$ and $\omega_2$, respectively, to the total LDOS, ${\tilde \rho ^{(e)}}({\mathbf{r}}) = {\rho ^{(e)}}({\mathbf{r}};{\omega _1})/\left[ {{\rho ^{(e)}}({\mathbf{r}};{\omega _1}) + {\rho ^{(m)}}({\mathbf{r}};{\omega _2})} \right]$ and ${\tilde \rho ^{(m)}}({\mathbf{r}}) = 1 - {\tilde \rho ^{(e)}}({\mathbf{r}})$, Fig.~\ref{spectr}(d).

Finally, intriguing results were reported in Ref.~\cite{S.Karaveli2013}. Authors managed to modulate the spontaneous emission spectrum of europium ions on the timescale less than the lifetime of the excited state. They placed Eu$^{3+}$ ions on the moving mirror, which was driven by a piezoelectric actuator powered with a sinusoidal voltage signal. In the experiment the distance between the ions layer and the metallic mirror was oscillating near the mean value $\approx$250~nm with amplitude $\approx$15~nm and 7.5~KHz frequency. While the total spontaneous emission rate of the excited state remained almost constant, the electric (magnetic) LDOS was substantially increased (decreased) or decreased (increased) in far or close positions of the mirror, respectively, Fig.~\ref{modulation}(a). In Ref.~\cite{ZiaNatComm2015} this kind of the LDOS modulation was also demonstrated for Er$^{3+}$ ions by using VO$_2$ phase-change material, Fig.~\ref{modulation}(b). While the lifetime of the excited state of erbium ions is several milliseconds, the ultrafast transition between the dielectric and metallic states of vanadium dioxide was reached on the scale of microseconds, Fig.~\ref{modulation}(c).

Overall, simple structures such as bulk dielectrics and planar structures provide a comprehensive basis for studying and controlling fluorescence properties of emitters with ED and MD transitions. Due to the presence of competition between several transitions from the same excited state with long lifetime emission spectrum of rare-earth ions can be modulated on the timescale of several microseconds and less by LDOS modulation which paves a new way to the development of on-chip optical communication.

\section{Magnetic Purcell factor in nanoantennas}
\subsection{Plasmonic nanoantennas}

\begin{figure*}
\centering
\includegraphics[width=0.99\textwidth]{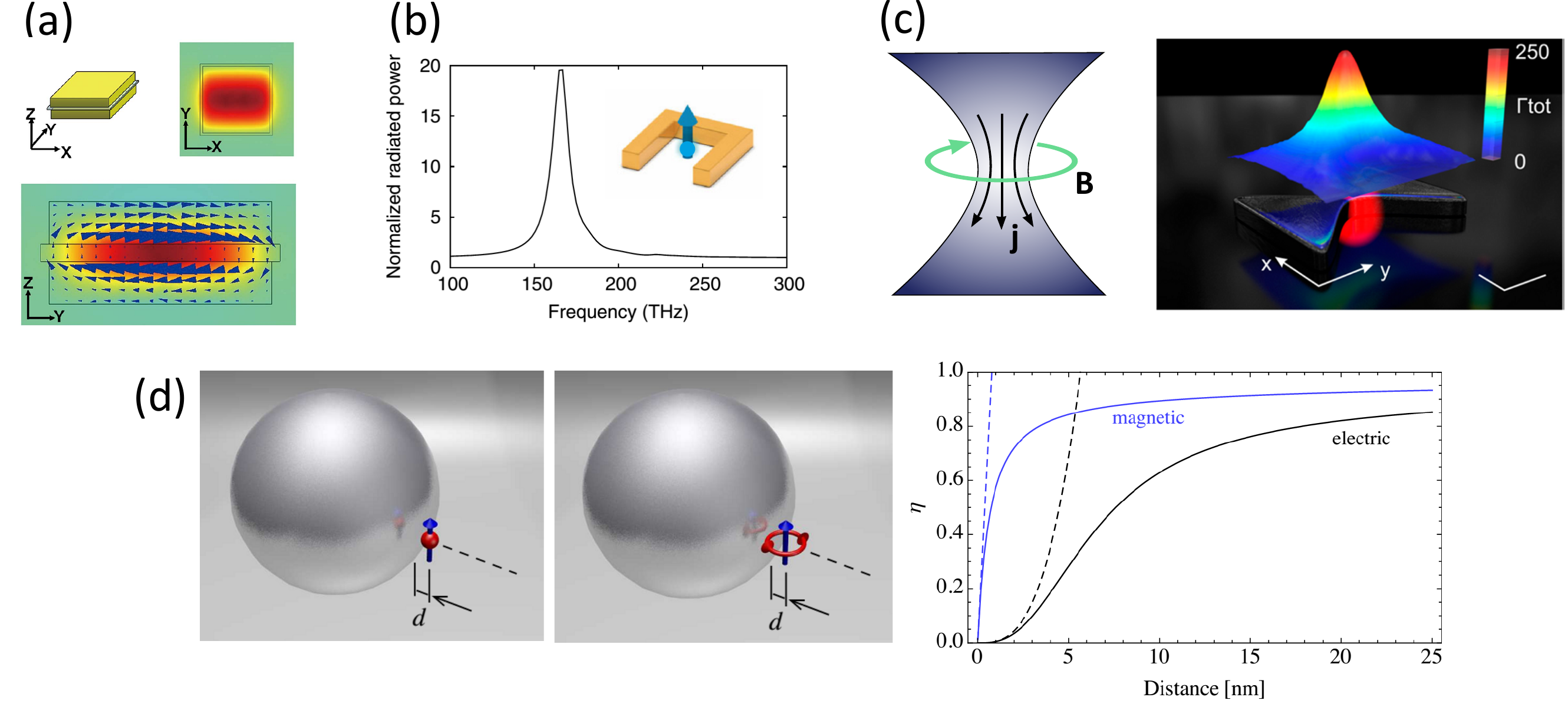}
\caption{Magnetic Purcell factor in plsamonic nanoantennas. (a)~Geometry and spatial electric field distribution of a magnetic mode of a patch plasmonic nanoantenna. Adapted from Ref.~\cite{TianhuaFeng2011}. (b)~Spectrum of the radiative enhancement of the MD spontaneous emission via split-ring resonator. Adapted from Ref.~\cite{Hein2013}. (c)~Geometry of a diabolo nanoantenna and the spatial map of the total decay rate enhancement for a MD emitter. Adapted from Ref.~\cite{Mivelle2015}. (d)~Schematic of an electric and magnetic dipole emitters located close to the surface of a metallic nanosphere and the radiation efficiency of electric and magnetic emitters as a function of distance between the emitter and the nanosphere. Adapted from Ref.~\cite{Chigrin2015}.}
\label{plasmonic}
\end{figure*}

Another approach towards enhanced spontaneous emission of MD sources relies on the use of \emph{optical nanoantennas}. In the recent years the outstanding progress in the development of nanoantennas enabling accelerated spontaneous emission has been witnessed. Usually, such antennas incorporate plasmonic nanoparticles~\cite{CarminatiReview, Pelton, Lakowicz2008, Brongersma_2010, Ramamurthy_2016}, in which the Purcell effect originates from the electric response of the nanostructure. However, enhancement of MD emission requires alternative nanoantennas design supporting magnetic resonances~\cite{Monticone2014, Shafiei2013, Verre2015}.

Historically, the first attempts to alternate the spontaneous emission of MD transitions in rare-earth ions via nanoantennas have been performed with the use of plasmonic nanoparticles~\cite{Deki_2003}. In this work the enhancing and quenching effects of silver (Ag) nanoparticles (of sizes less than 100~nm) on luminescence of Eu$^{3+}$ complexes in the solution phase have been experimentally investigated. It was shown that the luminescence intensity increases for small concentrations of Ag nanoparticles and then decreases upon addition of the nanoparticles. The authors claimed that the observed luminescence intensity is regarded as the result of the delicate balance between the particle size and their concentration, whereas the quenching is a result of relatively high dissipative losses of Ag particles in the visible range. Similarly, the enhancement of upconversion fluorescence of erbium ions in Er$^{3+}$: Ag-antimony glass nanocomposites caused by inclusions of nanosilver has been studied in Ref.~\cite{Karmakar_2009}.

Later, a \textit{single nanoantenna} enhanced MD emission was for the first time theoretically studied in Ref.~\cite{TianhuaFeng2011} for a hybrid metal-dielectric antenna, Fig.~\ref{plasmonic}(a). The nanostructure was formed by thin glass layer placed between two identical square gold patches. Currents induced in the two gold patches flowing in the opposite direction give rise to the magnetic dipole resonance of the structure (see Fig.~\ref{plasmonic}(a), the bottom picture). The total MD decay rate was theoretically demonstrated to be enhanced by a factor of 2000, whereas the radiative decay rate increases by a factor of 400. It has been shown, that the resonance wavelength of the nanoantenna can be continuously tuned in the range from 600~nm up to 2.2~$\mu$m via variation of the width of the gold patches. Moreover, the structure also exhibits low sensitivity to the position of a MD emitter.

The well-studied example of a metallic nanoantenna demonstrating optical magnetic response is the split-ring resonator~\cite{PendrySRR}. Magnetic response of such nanostructure is related to the electric current circulating along the ring. Enhanced magnetic response of such antennas, in particular, has facilitated the development of optical magnetic metamaterials~\cite{GiessenSRR, Linden2004}. Magnetic resonance of a split-ring resonator can also increase the Purcell factor of MD emitters, as was shown theoretically in Ref.~\cite{Hein2013}. Again, the total decay rate can be increased by orders of magnitude, while the radiative rate enhancement can be as large as 20 for the case of gold nanoantenna provided that the emitter is located in the antenna hot-spot (see Fig.~\ref{plasmonic}(b)). Such low quantum yield stems from efficient excitation of non-radiating dark modes, leading to large quenching of the spontaneous emission.

\begin{figure*}
\centering
\includegraphics[width=0.99\textwidth]{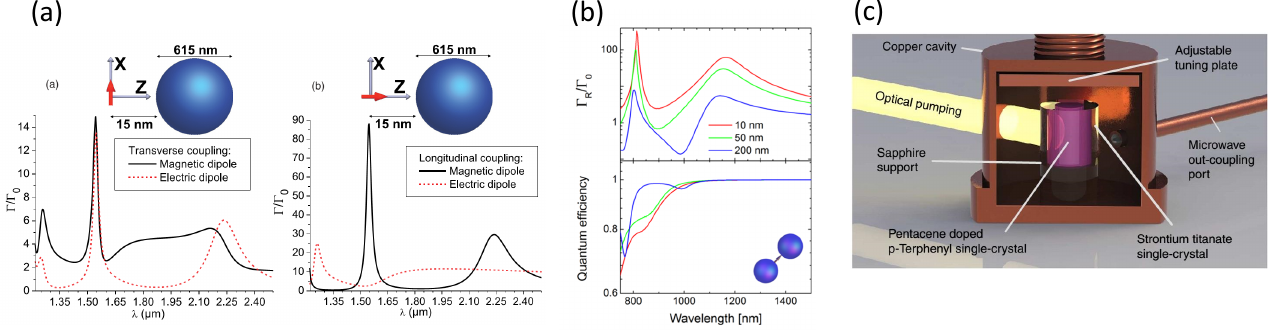}
\caption{Magnetic Purcell factor in dielectric nanoantennas. (a)~Total decay rates of an electric and magnetic dipole emitter coupled to a dielectric nanosphere. Adapted from Ref.~\cite{Rolly}. (b)~Enhancement of radiative decay rate and the corresponding quantum yield of a magnetic dipole emitter placed in a gap between two dielectric nanoparticles. Adapted from Ref.~\cite{Albella2013}. (c)~The design of resonant microwave cavity for room-temperature maser based on high-index dielectric material. Adapted from Ref.~\cite{Maser}.}
\label{dielectric}
\end{figure*}

An alternative geometry for the magnetic Purcell factor enhancement is represented by the \textit{diabolo nanoantenna}~\cite{Fischer, Park_2011, Mivelle2015}, as depicted in Fig.~\ref{plasmonic}(c). The difference between such nanoantenna and the well-known bow-tie geometry (see Refs.~\cite{Hecht_obz_12, HulstNP2011}) is that its two metallic patches are not isolated in its central region. This results in enhanced electric currents flowing through this gap, and thus enhanced magnetic field between the two patches~\cite{Fischer}. Numerical simulations of the nanodiabolo geometry revealed 2900-fold enhancement of the magnetic field at wavelength of 2540~nm, confined to a 40-by-40~nm region near the center of the nanoantenna. Such strong magnetic field enhancement leads to huge magnetic source--diabolo nanoantenna interaction. The unique capabilities of the diabolo nanoantennas for MD Purcell effect have been studied in Ref.~\cite{Mivelle2015}. The results of numerical simulations indicated that the total decay rate of a MD emitter can be increased by two orders of magnitude (see Fig.~\ref{plasmonic}(c)) with nearly unity quantum yield $\eta$.

It is known that the greater fraction of emission of an ED source placed on a nanometer distance from a plasmonic nanoparticle converts into heat via optical absorption~\cite{Anger2006, Vigoureux_2006, Castanie2010, Delga_2014}. This phenomenon, referred to as \textit{quenching}, makes extraction of light from quantum sources with the use of plasmonic nanostructures challenging. The problem of plasmonic quenching for MD emitters was addressed in Ref.~\cite{Chigrin2015}, where it was demonstrated that the non-radiative decay of a MD emitter placed near a plasmonic nanosphere is much weaker than that of an ED emitter (see Fig.~\ref{plasmonic}(d)). Specifically, the quantum yield of a MD source reaches 50\% at 1~nm distance (depicted as $d$) from the silver (Ag) nanosphere, whereas the magnetic Purcell factor reaches 5. At the same time, the quantum yield of an ED emitter at the same distance from the nanosphere is orders of magnitudes smaller.

The spontaneous emission rate from \textit{chiral molecules} can be also significantly altered by changing chirality of the environment. This was theoretically shown in Ref.~\cite{Klimov2012}, where the spontaneous emission rate of a chiral molecule placed near a chiral plasmonic sphere was investigated. The dramatic difference (up to 50 times) of spontaneous emission rates was observed for the two enantiomers. This effect may lead to promising applications for photoinduced separation of enantiomers of organic molecules.

To conclude the discussion of plasmonic nanoantennas, we would like to underline that, despite their unique opportunities for electric local field enhancement, plasmonic nanoantennas made from metals (gold, silver) have a number of disadvantages, including high dissipative losses, and the difficulty of achieving optical magnetic response by means of complex nanostructures such as rings of plasmonic nanoparticles~\cite{Simovski_2010, Shafiei2013, Rockstuhl_2013, Monticone2014}. Those limitations may be beaten with the use of all-dielectric nanoantennas, whose capabilities for enhancement of MD emission we address below.

\subsection{All-dielectric nanoantennas}
 
Today the vast majority of photonic structures exhibiting artificial magnetism in the visible range contain metallic elements, so that the strong dissipative losses have retarded their practical applications. To overcome this severe impediment and achieve the optical magnetic response in visible and near-IR, \textit{all-dielectric nanoantennas} based on high-index dielectric nanoparticles have been proposed. Under the magnetic dipole resonant condition, the polarization of the electric field is anti-parallel at opposite boundaries of the sphere, which gives rise to strong coupling of light to circular currents inside the sphere. In this section, we discuss the applications of all-dielectric nanoantennas for boosting the magnetic Purcell effect.
  
In 2010 it was theoretically shown that crystalline silicon (c-Si) nanoparticles may manifest the magnetic dipole resonance in the visible range~\cite{Evlyukhin_10_PRB}.
Then, the scattering properties of silicon nanoparticles have been studied in details~\cite{Garcia-Etxarri2011}. Shortly after, the concept of "\textit{magnetic light}" has been experimentally observed in the visible~\cite{Kuznetsov2012, Polman2013, Evlyukhin}, infrared~\cite{Shi2012}, and microwave~\cite{Geffrin2012} frequency ranges.
Such nanoparticles have attracted significant attention for their resonant behavior in recent years and thus have been widely employed for enhancing the light-matter interaction~\cite{Krasnok_2012, Krasnok_PU_2013, KrasnokLPR2015, ShcherbakovNL, Baranov16, Krasnok_APL_2016, Makarov_Nanoscale_2016, Bonod_NL_2016, Bonod_ACS_2016, PlasmaACS}.

\begin{figure*}
\centering
\includegraphics[width=0.99\textwidth]{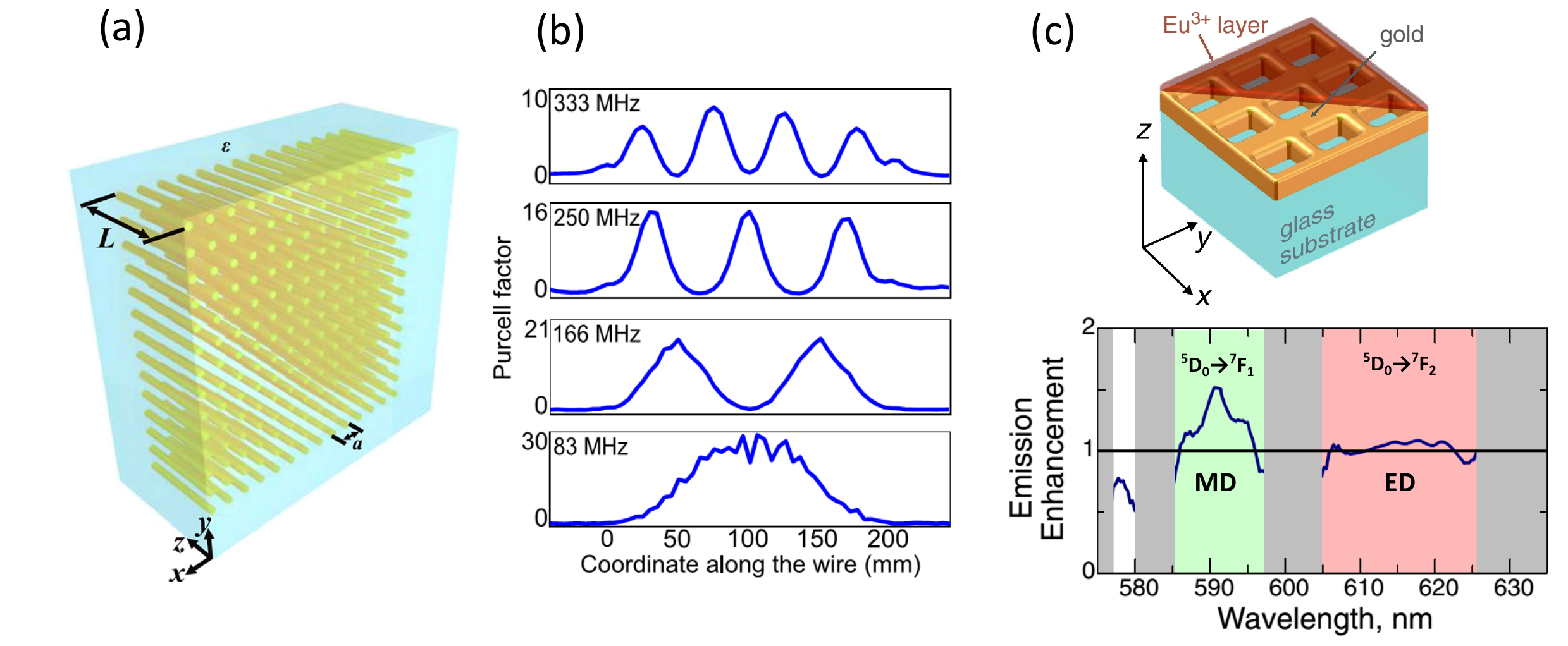}
\caption{Magnetic Purcell effect in metamaterials. (a)~Schematic of a finite size array of brass wires in dielectric matrix. (b)~Purcell factor as a function of the coordinate along the wires for four different frequencies. Adapted from Ref.~\cite{A.P.Slobozhanyuk2014}. (c)~Geometry of the resonant plasmonic nanostructure for enhancement of MD emission and the measured photoluminescence intensity from Eu$^{3+}$ ions. Adapted from Ref.~\cite{RabiaHussain2015}.}
\label{meta}
\end{figure*}

Magnetic modes of high-index nanoparticles also provide a natural way for enhancement of spontaneous emission from MD emitters due to their strong magnetic response~\cite{Aizpurua2012, Rolly, Zambrana-Puyalto2015, Krasnok_SR2015}. An important advantage of dielectric nanoantennas over plasmonic counterparts is their \textit{low dissipative losses}. For frequencies below the semiconductor bandgap, the optical absorption within semiconductor is negligible, so that the total and radiative decay rates become nearly equal. The general behavior of the Purcell factor for both ED and MD emitters enhanced by the Mie resonances of a dielectric nanoparticle is shown in Fig.~\ref{dielectric}(a). An important feature observed from this plot is that for longitudinal orientation of a point source with respect to the particle surface the emission of an electric or magnetic dipole is enhanced only by the respective resonance mode of the particle. At the same time, for transverse orientation of the source, the magnetic dipole couples to electric modes of the particle, and vice versa. 

A drawback of this approach is that the electric and magnetic hotspots of the Mie modes are concentrated within the particle, thus the region of the high magnetic LDOS is unavailable for positioning of a MD emitter. To overcome this, one may compose a dimer of resonant dielectric nanoantennas with the magnetic field hot-spot in the gap between the particles~\cite{Dimers1, Bakker2015, MaierNC_2015}. Enhancement of MD emission in dimers of silicon particles was theoretically predicted in Ref.~\cite{Albella2013}. More than 100-fold magnetic Purcell factor enhancement has been observed for transverse orientation of an emitter in a 10~nm gap between two silicon nanoparticles (see Fig.~\ref{dielectric}(b)). At the same time, the dimer nanostructure demonstrates nearly unity quantum yield in the broad range of emitter wavelengths, as one can see in Fig.~\ref{dielectric}(b). Local enhancement of the magnetic field, related to the accelerated emission of MD transitions, was also studied for dimers of cubic dielectric nanoparticles~\cite{Boudarham2014}, which may be more accessible from the fabrication point of view. 
Hollow silicon nanodisks have been also analyzed theoretically in the context of enhancing MD emission. Due to the presence of a hole in the center of the nanodisk, the region of enhanced magnetic field becomes accessible for positioning of rare-earth ions, what allows achieving the radiative magnetic Purcell factor as high as 300~\cite{nanodisk2016}.

Enhancement of magnetic Purcell factor via dielectric resonators in the microwave was employed in Ref.~\cite{Maser} for the development of a maser operating at room temperature. In this work, a dielectric cylinder (SrTiO$_3$) was used as a maser cavity, Fig.~\ref{dielectric}(c). Large permittivity of this material in the microwave range ($\sim 300$) enables very high magnetic Purcell factor ($\sim 3.6 \times 10^7$), allowing for reducing the maser threshold up to 2~W at room temperature.

\section{Magnetic Purcell effect in metamaterials}

Nanostructured metamaterials offer an alternative approach for enhancement of MD spontaneous emission. In contrast to the case of ED emitters, for which enhanced spontaneous emission was reported in a lot of studies for various configurations~\cite{Narimanov, Shalaev, Decker, Acherman}, to the best of our knowledge only two works have demonstrated enhanced spontaneous emission of MD emitters in a nanostructured environment.

The so-called wire metamaterials (see~\cite{Wire1} and references therein) are known for their infinitely extended TEM modes in isofrequency contours, which are responsible for high LDOS and consequently huge spontaneous emission enhancement, similarly to other metamaterials with hyperbolic dispersion~\cite{Poddubny2013}. Because of the strong spatial dispersion in metallic wire metamaterials~\cite{Belov2003} an accurate theory of spontaneous emission in such structures requires accounting for the discreteness of the structure and the finite value of wires permittivity. Such theory was developed in Ref.~\cite{Wire2} for both ED and MD types of emitter. It was shown that due to the coupling of the in-plane oriented (perpendicular to the wires) MD source placed in the center of the unit cell to TEM modes of the wire metamaterial, strong enhancement (up to 100) can be achieved for large values of permittivity of the wires. Moreover, the value of Purcell factor strongly depends on the position of the source and it increases up to several thousands for the sources located near the wires. Enhancement of the MD Purcell factor has also been demonstrated theoretically for a dielectric wire medium exhibiting magnetic hyperbolicity~\cite{wire16} and for a hexagonal array of composite Ag/Al nanorods~\cite{Cui:16}.

The experimental evidence of MD spontaneous emission enhancement provided by a wire metamaterial has been reported in Ref.~\cite{A.P.Slobozhanyuk2014}. The authors have measured magnetic Purcell factor for the structure composed of $14\times14$ brass wires placed in the reservoir with distilled water, which is dielectric in microwave frequency range (see Fig.~\ref{meta}(a)). A subwavelength loop antenna was employed as a MD source. The design of the experiment allowed measurements of the Purcell factor placed above the metamaterial slab as a function of the source position along the wires and frequency, as shown in Fig.~\ref{meta}(a). The measurements revealed the strong spontaneous emission enhancement at the frequencies corresponding to the subwavelength-volume Fabry-P$\acute{\mathrm{e}}$rot modes of the wire medium resonator (Fig.~\ref{meta}(b)). The highest value of Purcell factor $\approx$30 was measured for the lowest order mode when the source was placed in the middle of the wire where the mode profile has the maximum of the magnetic field.

Wire metamaterials represent the nonresonant type of metamaterials in contrast to resonant metamaterials, whose properties mostly stem from the resonances of constitutive elements. The magnetic Purcell factor in a resonant system was firstly demonstrated in the Ref.~\cite{RabiaHussain2015}. Authors employed a thin layer of Eu$^{3+}$ ions that was placed above an array of holes in a gold film (Fig.~\ref{meta}(c)). Such structure is characterized by plasmonic resonance in the spectral region of the MD transition of Eu$^{3+}$ ($\approx$590~nm). Emission spectra comparison of the europium layer with the gold metasurface and without one revealed that the spontaneous emission rate of MD transition increases for about 50\%. Time-resolved measurements of MD spontaneous emission enhanced by plasmonic metasurfaces have been performed in Ref.~\cite{Choi}, revealing 3.5-fold acceleration of MD emission from Er$^{3+}$ ions at 1.53~$\mu$m.

\section{Conclusion and Outlook}

To conclude, we have covered the recent advances in the area of enhanced spontaneous emission from magnetic dipole emitters. Tailoring of the emission lifetime of magnetic dipole transitions is gaining the considerable interest due to the latest development of nanofabrication methods. Ions of rare-earth metals have proven to be versatile emission sources in visible and near-IR, that demonstrate magnetic dipole emission at specific wavelengths. Modification of the spontaneous emission rate of magnetic dipole transitions has been successfully demonstrated in experiments with various rare-earth ions with the use of planar structures and nanostructured metamaterials. The results of these experiments have led to the development of nanoscale probes for direct measurements of the magnetic local density of states. At the same time, while from the theoretical standpoint plasmonic and all-dielectric nanoantennas represent the most promising platform for enhancement of magnetic dipole spontaneous emission, the experimental evidences of such enhancement have not been yet reported. Nevertheless, we expect that the current progress in nanofabrication will allow for observation of nanoantenna assisted enhanced magnetic dipole emission in the nearest future.

\begin{acknowledgments}
The authors are thankful to Grigory~Ptitsyn, Yali~Sun, Ivan Iorsh, Alexander Poddubny, and Pavel Ginzburg for stimulating discussion and critical comments to our manuscript.
\end{acknowledgments}

\bibliographystyle{apsrev4-1}
\bibliography{bibliography}

\end{document}